\documentclass[aps,pra,twocolumn,showpacs,superscriptaddress]{revtex4}
\usepackage{graphicx}
\usepackage{amsmath}
\usepackage{amsfonts}
\usepackage{color}
\usepackage{bm}
\newcommand{\ket}[1]{\ensuremath{\left|{#1}\right\rangle}}
\newcommand{\bra}[1]{\ensuremath{\left\langle{#1}\right |}}

\newcommand{\oper}[1]{\mathbf{\mathsf{#1}}}

\newcommand{\sinc}{\ensuremath{{\mathrm{sinc}}}}

\newcommand{\beq}{\begin{equation}}
\newcommand{\eeq}{  \end{equation}}
\newcommand{\bea}{\begin{eqnarray}}
\newcommand{\eea}{  \end{eqnarray}}
\newcommand{\bit}{\begin{itemize}}
\newcommand{\eit}{  \end{itemize}}

\begin{document}


\title{An Entropic Einstein-Podolsky-Rosen Criterion}

\author{S. P. Walborn}
\affiliation{Instituto de F\'{\i}sica, Universidade Federal do Rio
de Janeiro, Caixa Postal 68528, Rio de Janeiro, RJ 21941-972,
Brazil}
\author{A. Salles}
 \affiliation{Instituto de F\'{\i}sica,
Universidade Federal do Rio de Janeiro, Caixa Postal 68528, Rio de
Janeiro, RJ 21941-972, Brazil}

\author{R. M. Gomes}
 \affiliation{Instituto de F\'{\i}sica,
Universidade Federal do Rio de Janeiro, Caixa Postal 68528, Rio de
Janeiro, RJ 21941-972, Brazil}
\affiliation{Instituto de F\'\i sica, Universidade Federal de Goi\'as, Goi\^ania GO 74.001-970, Brazil}

\author{F. Toscano}
 \affiliation{Instituto de F\'{\i}sica, Universidade Federal do Rio
              de Janeiro, Caixa Postal 68528, Rio de Janeiro, RJ 21941-972,
              Brazil}
\author{P. H. Souto Ribeiro}
 \affiliation{Instituto de F\'{\i}sica,
Universidade Federal do Rio de Janeiro, Caixa Postal 68528, Rio de
Janeiro, RJ 21941-972, Brazil}

\date{\today}

\begin{abstract}
We propose an EPR inequality based on an entropic uncertainty relation for complementary continuous variable observables.  This inequality is more sensitive than the previously established EPR inequality based on inferred variances, and opens up the possibility of EPR tests of quantum nonlocality in a wider variety of quantum states.  We experimentally test the inequality using spatially entangled photons.  For a particular quantum state, our experimental results show a violation of the entropic EPR inequality, while the variance EPR inequality is not violated.       
  \end{abstract}


\maketitle
\par
In 1935, Einstein, Podolsky and Rosen published the famous ``EPR" paper, which initiated the study and investigation of quantum entanglement and quantum non-locality \cite{epr35}.  EPR argued that the perfect correlations present in a particular pair of spatially separated quantum systems could be used to simultaneously ascribe well-defined values to complementary variables.  Reasoning that any reasonable theory must be local and attribute ``elements of reality" to physical quantities, EPR concluded that quantum theory, which does not do so, must be incomplete.  The EPR paradox, as it is now known, ignited the ongoing discussion concerning locality, realism and quantum entanglement, which has intensified recently due to applications in quantum information science.  In 1964, John S. Bell used the discrete version of the EPR paradox to derive the Bell inequality, thus leading the way to experimentally test the predictions of quantum mechanics itself against those of local realism \cite{bell87}.   Early experiments by Aspect and collaborators \cite{aspect82a}, and more recent experiments since then \cite{weihs98,baas08}, indicate that  quantum mechanics cannot be described by EPR's local realism.    
\par
In the continuous variable regime, it has been shown that, for non-commuting variables satisfying $[\oper{X},\oper{P}]=i$, local realism is in conflict with the completeness of quantum mechanics when the inequality \cite{reid88,reid89}    
\begin{equation}
\Delta^2_{\mathrm{min}}(X_A) \Delta^2_{\mathrm{min}}(P_A) \geq \frac{1}{4}
\label{eq:reid}
\end{equation}     
is violated.  Here $\Delta^2_{\mathrm{min}}(X_A)$ is the minimum uncertainty in inferring property $X_A$ of system $A$ given measurement of property $X_B$ on system $B$, and is given by
$\Delta^2_{\mathrm{min}}(X_A) = \int dx_B \mathcal{P}(x_B) \Delta^2(x_A|x_B)$, where $\Delta^2(x_A|x_B)$ is the variance of the conditional probability distribution $\mathcal{P}(x_A|x_B)$ \cite{reid08}.
Throughout this letter we use uppercase letters to represent the random variables, and lowercase letters to depict their possible values, so that the probability density functions are $\mathcal{P}(x)$.  The EPR criterion \eqref{eq:reid} considers a situation nearly identical to the original EPR gedanken experiment, with the advantage that it applies to the more realistic scenario of non-perfect correlations.     
It was first violated for quadrature measurements of intense beams \cite{ou92,silberhorn01,schori02,bowen03,marino09}, and subsequently for ``spin" variables of light beams \cite{bowen02} and  spatial degrees of freedom of entangled photons \cite{howell04,dangelo04,tasca09}. 
\par
In this letter, we introduce an EPR criterion that is based on  the conditional Shannon entropy. We demonstrate both theoretically and experimentally that it is more sensitive than the variance EPR criterion for non-gaussian states. For gaussian states, we show that it is equivalent to the variance criterion \eqref{eq:reid}.
\par
Let us briefly summarize the logical steps necessary to obtain the  EPR criterion \eqref{eq:reid}. A more detailed discussion is available in Refs. \cite{reid89}, and in a recent review paper \cite{reid08}.  Consider that an $x$ measurement on system $B$ with result $x_B$ indicates that the probability of obtaining result $x_A$ on system $A$ is given by the conditional probability distribution $\mathcal{P}(x_A|x_B)$, and similarly for the variable $p$, $\mathcal{P}(p_A|p_B)$. The assumption of local realism implies that the conditional probability distributions for $x$ and $p$ measurements correspond to simultaneous \emph{elements of reality}, since measurement at $B$ can in no way affect the properties of $A$.      
If the local realistic description is consistent with quantum mechanics, then it should reproduce the predictions of quantum theory.   One can consider then the quantum mechanical uncertainty relation 
\begin{equation}
\Delta^2({X_A})\Delta^2({P_A}) \geq \frac{1}{4}.
\label{eq:heisenberg}
\end{equation}
If  $\mathcal{P}(x_A|x_B)$ and  $\mathcal{P}(p_A|p_B)$ are simultaneous elements of reality, then the variances of these distributions should satisfy the uncertainty relation \eqref{eq:heisenberg}, which leads to the EPR criterion \eqref{eq:reid}.  It has been pointed out by Cavalcanti and Reid that one can construct an EPR criterion from any quantum mechanical uncertainty relation \cite{cavalcanti07}.      
\par
Let us now propose an entropic EPR criterion.
The $x$ and $p$ distributions of a quantum system must  satisfy the entropic uncertainty relation \cite{bialynicki75}:
 \begin{equation}
 h(X) + h(P) \geq \ln \pi e,
 \label{eq:entunc}
 \end{equation}
 where $h(R)=-\int dr \mathcal{P}(r) \ln \mathcal{P}(r)$ is the differential Shannon entropy.       
 Following the same arguments as above, with the assumption of local realism, the conditional probability distributions must satisfy the entropic uncertainty relation:  $h(X_A|X_B=x_B) + h(P_A|P_B=p_B) \geq \ln \pi e$, where we define $h(R_A|R_B=r_B)=-\int dr_A \mathcal{P}(r_A|r_B) \ln \mathcal{P}(r_A|r_B)$. 
Multiplying this inequality by $\mathcal{P}(x_B)\mathcal{P}(p_B)$ and integrating over $x_B$ and $p_B$ gives an entropic EPR criterion:
 \begin{equation}
 h(X_A|X_B) + h(P_A|P_B) \geq \ln \pi e, 
 \label{eq:entEPR}
 \end{equation}
where the conditional entropy is defined as \cite{cover}
 \begin{equation}
 h(R_A|R_B) =-\int dr_B \mathcal{P}(r_B)h(R_A|R_B=r_B).
 \label{eq:condent}
 \end{equation}
Violation of inequality \eqref{eq:entEPR} indicates a physical situation for which local realism is inconsistent with the completeness of quantum mechanics.  The entropy of a probability distribution with variance $\Delta^2(r_A|r_B)$ is upper-bounded by   
$\ln[2 \pi e \Delta^2(r_A|r_B)]/2$ \cite{cover}.  Thus, we have as an upper bound: 
\begin{equation}
 \ln[2 \pi e \Delta^2(x_A|x_B)\Delta^2(p_A|p_B)] \geq h(X_A|X_B) + h(P_A|P_B).
 \label{eq:bound}
\end{equation}
The upper limit is reached when both conditional probabilities are Gaussian \cite{cover}, and inequality \eqref{eq:entEPR} is equivalent to the variance-product EPR criterion \eqref{eq:reid}.  Thus, inequality \eqref{eq:bound} shows that the entropic inequality \eqref{eq:entEPR} is in general more sensitive than the variance inequality \eqref{eq:reid}.      
This indicates, as we will now show, that the entropic EPR criterion \eqref{eq:entEPR} is capable of recognizing an EPR paradox in a wider variety of continuous variable quantum states.   
\begin{figure}
\begin{center}
\includegraphics[width=8cm]{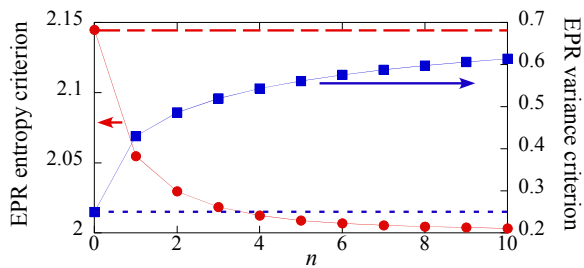}
\end{center}
\caption{\label{fig:plot_epr} (Color online) The EPR entropy (red circles) and variance criteria (blue squares) as a function of $n$ for wave function \eqref{eq:phi}.  The long-dashed line is the lower limit for the entropic criterion, while the short-dashed line is the limit for the variance criterion.  The EPR paradox is identified below these limits.}
\end{figure}

\par
Consider now a bipartite quantum system described by the wave function
\begin{equation}
\phi(x_A,x_B) = C_{n}\mathcal{H}_{n}(x_A+x_B)\,e^{-\frac{(x_A+x_B)^2}{2}}e^{-\frac{(x_A-x_B)^2}{2}},  
\label{eq:phi}
\end{equation} 
where $\mathcal{H}_{n}$ is the $n^{\mathrm{th}}$-order Hermite polynomial and $C_n$ a normalization constant.  This pure state is entangled for $n>0$.  FIG. \ref{fig:plot_epr} shows the EPR variance criterion \eqref{eq:reid} and EPR entropy criterion as a function of $n$, as well as the lower limits for both criteria.  The entropic criterion identifies an EPR paradox for all $n>0$, while the variance criterion does not. 
\par
\begin{figure}
\begin{center}
\includegraphics[width=8cm]{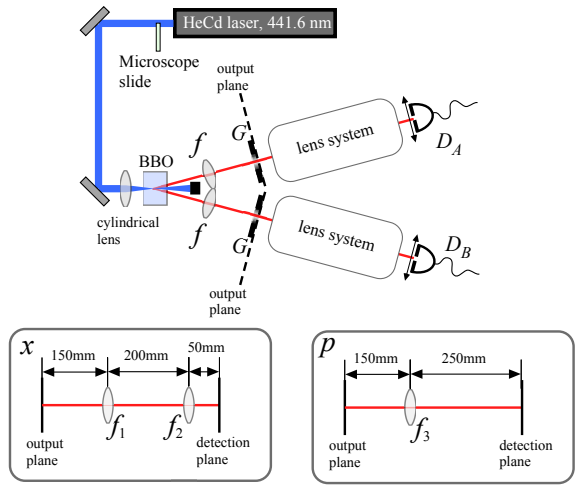}
\end{center}
\caption{\label{fig:setup} (Color online) Experimental setup. SPDC in a BBO crystal produces spatially entangled photon pairs.  The lenses $f=100$mm are used to map the momentum distribution at the crystal face onto the output planes.  $G$ are transmission masks with a gaussian intensity profile.  The lens systems used for $x$ and $p$ measurements are shown in the boxes at the bottom.  Here $f_1=150$mm, $f_2=50$mm and $f_3=250$mm.  Detectors $D_A$ and $D_B$ are equipped with $40$nm FWHM interference filters, and are scanned in the vertical direction.}
\end{figure}

\par
We experimentally tested the entropic EPR criterion for a pair of spatially entangled photons.   
FIG. \ref{fig:setup} shows the experimental setup.  A HeCd laser is used to pump a $10$mm long type-II BBO crystal, producing photons centered around $\lambda=884$nm through spontaneous parametric down conversion (SPDC). The quantum state of the down-converted photons at the crystal is given to good approximation by \cite{monken98,walborn03a}   
 \begin{equation}
\label{eq:state}
\ket{\psi}=\iint dp_A dp_B v(p_A+p_B)\, s(p_A-p_B) \ket{p_A} \ket{p_B}, 
\end{equation}
where we consider only one spatial dimension for simplicity. 
Here $v(p)$ is the angular spectrum of the pump beam, and $p_A$, $p_B$ are the transverse wave vectors of the down-converted photons $A$ and $B$, respectively.  The  function $s(p) \propto 4K \sinc(L p^2/4K)$, where $L$ is the length of the BBO crystal and $K$ is the wave number of the pump beam.  
A number of steps were taken to engineer the wave function $\bra{x_A,x_B}\psi\rangle$ corresponding to the state \eqref{eq:state}, so that it was similar to that of Eq. \eqref{eq:phi} with $n=1$.
The pump laser is prepared in a Hermite-Gaussian mode using a glass microscope slide aligned to introduce a $\pi$ phase shift between the two halves of the Gaussian pump beam as in Ref. \cite{nogueira04}. The Hermite-Gauss profile of the pump beam is then passed on to the spatial profile of the down-converted photons \cite{monken98}. A set of Gaussian profile transmission masks $G$, placed in the focal plane of lenses ($f=100$mm), are used to filter the oscillatory tails of the $\sinc$ function.  This results in a two-photon wave function approximately described by 
\begin{equation}
\psi(x_A,x_B) = \frac{(x_A+x_B)}{\sqrt{\pi \sigma_- \sigma_+^3}}e^{-\frac{(x_A+x_B)^2}{4\sigma_+^2}}e^{-\frac{(x_A-x_B)^2}{4\sigma_-^2}}.  
\label{eq:psi}
\end{equation} 
This state is entangled for all values of $\sigma_\pm$.  The variance EPR inequality \eqref{eq:reid} is violated for $(\sigma_\mp/\sigma_\pm) \geq 6.615$.  The entropic EPR inequality, on the other hand, is violated for all values of $\sigma_\pm$.  Thus, the state described by wave function \eqref{eq:psi} should always manifest the EPR paradox. 
In our setup the pump beam was focused at the crystal face, so that $\sigma_+$ and $\sigma_-$ were of the same order of magnitude.  We measured $\sigma^2_+\approx 0.566$mm$^2$ and $\sigma^2_-\approx 0.240$mm$^2$ at the output planes.      
\par
\begin{figure}
\begin{center}
\includegraphics[width=8cm]{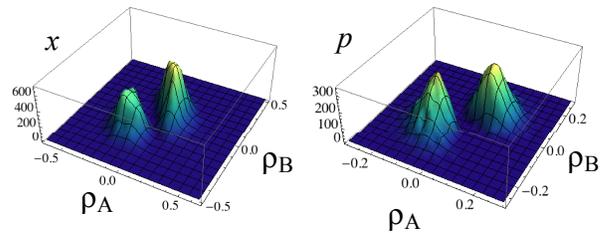}
\end{center}
\caption{\label{fig:results} Coincidence counts for $x$ and $p$ measurements used to calculate the probability distributions $\mathcal{P}(x_A,x_B)$ and $\mathcal{P}(p_A,p_B)$.}
\end{figure}

  An imaging lens system using two lenses ($f_1$ and $f_2$ in inset of FIG. \ref{fig:setup}) was used to measure the near field ($x$ variable), so that the output plane of the source was imaged at the detection plane. The far-field ($p$ variable) was measured by scanning in the focal plane of a second lens system ($f_3$ in inset of FIG. \ref{fig:setup}).  The lenses were mounted in detachable magnetic mounts, so that they could be easily switched in and out of the setup while maintaining alignment.  The detectors were scanned in discrete steps $z_{step}$ and coincidence counts were registered, resulting in two 2D tables of coincidence measurements  $C_{xx}(z_A,z_B)$ and $C_{pp}(z_A,z_B)$,  where $z_A$ and $z_B$ are the transverse positions of detectors $D_A$ and $D_B$, respectively.  The discrete probability distributions were then obtained by $\mathcal{P}_{rr}(z_A,z_B)=C_{rr}(z_A,z_B)/\sum_{z_A,z_B} C_{rr}(z_A,z_B)$, where $r=x,p$.  The coincidence measurements are shown in FIG. \ref{fig:results}.     
\par
We first tested the variance-product EPR inequality \eqref{eq:reid}, using the experimental data to compute $\Delta^2_{\mathrm{min}}(R_i) = \gamma_r^2\Delta^2_{\mathrm{min}}(Z_i)$, where $\Delta^2_{\mathrm{min}}(Z_i) = \sum_{z_j}\mathcal{P}(z_j)\Delta^2(z_i|z_j)$ for $i,j=A,B$.  Here $\Delta^2(z_i|z_j)$ is the variance in $z_i$ given result $z_j$ and $\gamma_r$ is the scaling factor, used to relate the detector positions $z$ to the $x$ and $p$ variables \cite{tasca09}.  Explicitely, $\gamma_x=f_2/f_1$, due to the magnification factor of the lens system, and $\gamma_p=2\pi/f_3\lambda$ for $p$ measurements.  We obtained $\Delta^2_{\mathrm{min}}(X_A)=0.14\pm 0.02$mm$^2$, $\Delta^2_{\mathrm{min}}(P_A) = 3.1\pm0.2$mm$^{-2}$, $\Delta^2_{\mathrm{min}}(X_B)=0.15\pm0.02$ mm$^2$, and $\Delta^2_{\mathrm{min}}(P_B)=3.4\pm0.2$mm$^{-2}$.  In all results reported here and below, the uncertainty in the experimental data was obtained by error propagation of the Poissonian count statistics.   The variance EPR inequality \eqref{eq:reid} gives  
\begin{align}
\Delta^2_{\mathrm{min}}(X_A) \Delta^2_{\mathrm{min}}(P_A) & = 0.44 \pm0.01 > \frac{1}{4}  \nonumber \\
\Delta^2_{\mathrm{min}}(X_B) \Delta^2_{\mathrm{min}}(P_B) & = 0.51 \pm0.01 > \frac{1}{4}.   
\end{align}    
Thus, the variance inequality is satisfied, and we cannot identify a conflict between the completeness of quantum mechanics and local realism. 

\begin{table}
 \begin{ruledtabular}
 \begin{tabular}{cl|cc}
$h(X_A)$ &$0.56\pm0.01\hspace{5mm}$  & $h(P_A)$ &$2.35 \pm0.01$  \\
$h(X_B)$ &$0.58\pm0.01$  & $h(P_B)$ &$2.37 \pm 0.01$ \\
$h(X_A,X_B)$ &$0.73\pm0.02$  & $h(P_A,P_B)$ & $4.17\pm0.03$  \\
\end{tabular}
 \end{ruledtabular}
 \caption{\label{tab:1}  Summary of experimental results.}
\end{table}

\par
Next we tested the entropic EPR inequality \eqref{eq:entEPR}.  The discrete entropies of the coincidence count distributions were calculated using  $H(Z)=-\sum_{z} \mathcal{P}(z) \ln \mathcal{P}(z)$, and $H(Z_A,Z_B)=\sum_{z_A,z_B} \mathcal{P}(z_A,z_B) \ln \mathcal{P}(z_A,z_B)$.  
The differential entropies $h(Z_A,Z_B)$ and $h(Z)$ of the continuous variables can be calculated from the discrete entropies $H(Z_A,Z_B)$ and $H(Z)$: 
\begin{subequations}
\label{eq:scale}
\begin{align}
h(Z_A,Z_B) & \approx H(Z_A,Z_B)+\ln(z_{\rm{step}}^2),  \\
 h(Z) & \approx H(Z)+\ln(z_{\rm{step}}). 
 \end{align}
 \end{subequations}
 Here $z_{\rm{step}}$ appears due to the discretization of the continuous distribution \cite{cover}.  For $x$ measurements $z_{\rm{step}}=0.02$mm, while $z_{\rm{step}}=0.05$mm for $p$ measurements.  
Finally, the entropy of the probability distributions for the $R=X,P$ variables are calculated from the experimental data using $h(R)=h(Z)+\ln \gamma_r$ and  $h(R_A,R_B)=h(Z_A,Z_B)+\ln \gamma_r^2$.       
Using these experimental results, the conditional differential entropies of the corresponding continuous probability distributions can be  calculated using $h(R_i|R_j)=h(R_i,R_j)-h(R_j)$, and are summarized in table \ref{tab:1}.    
Using this procedure, we calculated 
\begin{align}
h(X_A|X_B) + h(P_A|P_B) & = 1.94\pm0.04,  \\
h(X_B|X_A) + h(P_B|P_A) & = 1.99\pm0.04. 
\end{align}
Both of these equations are less than $\ln \pi e \approx 2.145$ by more than 3 standard deviations, indicating violation of inequality \eqref{eq:entEPR}.  Using the wavefunction \eqref{eq:psi}, the sum of conditional entropies was calculated to be $1.91$, showing considerably good agreement between theory and experiment.    
Thus, the EPR non-locality of the wave-function \eqref{eq:psi}, which is not identified under the variance criterion \eqref{eq:reid}, is revealed through test of the entropic EPR criterion \eqref{eq:entEPR}. 
\par
 
\par
Let us now briefly mention several applications of the entropic EPR criterion \eqref{eq:entEPR} in the context of quantum information science. It is known that violation of Bell's inequalities is fundamentally related to security bounds in entanglement-based quantum key distribution with qubits \cite{gisin02}.       
Concerning quantum key distribution with entangled continuous variable states \cite{reid00}, violation of the entropic EPR criterion \eqref{eq:entEPR} is sufficient to guarantee a secret key rate $\Delta I$.  In Ref. \cite{grosshans04}, it was shown that a lower bound for the secret key rate is given by 
\begin{equation}
\Delta I \geq \ln \pi e - h(X_B|X_A) - h(P_B|P_A). 
\end{equation}
It thus follows directly that the entropic EPR criterion \eqref{eq:entEPR} must be violated to achieve a non-zero key rate ($\Delta I > 0$).  
\par
An interesting question pertains to what states violate an EPR criterion.  It has been shown that all bipartite pure states violate some type of (discrete variable) Bell inequality, and thus, all pure, entangled bipartite states are Bell non-local \cite{gisin91,gisin92}.  It is interesting as to whether the same can be said for EPR criteria of continuous variables.  
Here we have made a step towards the affirmative answer to this question by opening up the possibility of violating a locality criterion for a greater number of non-Gaussian states. We note that any entropic uncertainty relation can be used to establish an entropic EPR criteria.  For example, application of the uncertainty relation for the R\'enyi \cite{bialynicki06} or Tsallis entropies \cite{wilk08} should lead to a family of entropic EPR inequalities.  We leave the careful investigation of this question to future work.    
\par
We have proposed an EPR criterion based on the Shannon entropy of conditional measurements, and tested it experimentally using the spatial degrees of freedom of photon pairs.  We obtain a clear violation of the entropic EPR inequality, while the variance EPR inequality is not violated.   Our theoretical and experimental results show that the entropic EPR inequality allows for the observation of EPR non-locality in a greater variety of states than the criterion based on inferred variances, and should allow for further investigation and characterization of tests of EPR non-locality.         
\begin{acknowledgements}
We thank R. L. Matos Filho and D. S. Tasca for helpful discussions.  Financial support was provided by Brazilian agencies CNPq, CAPES,
FAPERJ, FUJB, and the Instituto Nacional de Ci\^encia e Tecnologia - Informa\c{c}\~ao Qu\^antica.
\end{acknowledgements}


\begin{thebibliography}{30}
\expandafter\ifx\csname natexlab\endcsname\relax\def\natexlab#1{#1}\fi
\expandafter\ifx\csname bibnamefont\endcsname\relax
  \def\bibnamefont#1{#1}\fi
\expandafter\ifx\csname bibfnamefont\endcsname\relax
  \def\bibfnamefont#1{#1}\fi
\expandafter\ifx\csname citenamefont\endcsname\relax
  \def\citenamefont#1{#1}\fi
\expandafter\ifx\csname url\endcsname\relax
  \def\url#1{\texttt{#1}}\fi
\expandafter\ifx\csname urlprefix\endcsname\relax\def\urlprefix{URL }\fi
\providecommand{\bibinfo}[2]{#2}
\providecommand{\eprint}[2][]{\url{#2}}

\bibitem[{\citenamefont{Einstein et~al.}(1935)\citenamefont{Einstein, Podolsky,
  and Rosen}}]{epr35}
\bibinfo{author}{\bibfnamefont{A.}~\bibnamefont{Einstein}},
  \bibinfo{author}{\bibfnamefont{D.}~\bibnamefont{Podolsky}}, \bibnamefont{and}
  \bibinfo{author}{\bibfnamefont{N.}~\bibnamefont{Rosen}},
  \bibinfo{journal}{Phys. Rev.} \textbf{\bibinfo{volume}{47}},
  \bibinfo{pages}{777} (\bibinfo{year}{1935}).

\bibitem[{\citenamefont{Bell}(1987)}]{bell87}
\bibinfo{author}{\bibfnamefont{J.~S.} \bibnamefont{Bell}},
  \emph{\bibinfo{title}{Speakable and Unspeakable in Quantum Mechanics}}
  (\bibinfo{publisher}{Cambridge University Press}, \bibinfo{address}{New
  York}, \bibinfo{year}{1987}).

\bibitem[{\citenamefont{Aspect et~al.}(1982)\citenamefont{Aspect, Grangier, and
  Roger}}]{aspect82a}
\bibinfo{author}{\bibfnamefont{A.}~\bibnamefont{Aspect}},
  \bibinfo{author}{\bibfnamefont{P.}~\bibnamefont{Grangier}}, \bibnamefont{and}
  \bibinfo{author}{\bibfnamefont{G.}~\bibnamefont{Roger}},
  \bibinfo{journal}{Phys. Rev. Lett.} \textbf{\bibinfo{volume}{49}},
  \bibinfo{pages}{91} (\bibinfo{year}{1982}).

\bibitem[{\citenamefont{Weihs et~al.}(1998)\citenamefont{Weihs, Jennewein,
  Simon, Weinfurter, and Zeilinger}}]{weihs98}
\bibinfo{author}{\bibfnamefont{G.}~\bibnamefont{Weihs}},
  \bibinfo{author}{\bibfnamefont{T.}~\bibnamefont{Jennewein}},
  \bibinfo{author}{\bibfnamefont{C.}~\bibnamefont{Simon}},
  \bibinfo{author}{\bibfnamefont{H.}~\bibnamefont{Weinfurter}},
  \bibnamefont{and}
  \bibinfo{author}{\bibfnamefont{A.}~\bibnamefont{Zeilinger}},
  \bibinfo{journal}{Phys. Rev. Lett.} \textbf{\bibinfo{volume}{81}},
  \bibinfo{pages}{5039} (\bibinfo{year}{1998}).

\bibitem[{\citenamefont{Baas et~al.}(2008)\citenamefont{Baas, Branciard, Gisin,
  Zbinden, and Salart}}]{baas08}
\bibinfo{author}{\bibfnamefont{A.}~\bibnamefont{Baas}},
  \bibinfo{author}{\bibfnamefont{C.}~\bibnamefont{Branciard}},
  \bibinfo{author}{\bibfnamefont{N.}~\bibnamefont{Gisin}},
  \bibinfo{author}{\bibfnamefont{H.}~\bibnamefont{Zbinden}}, \bibnamefont{and}
  \bibinfo{author}{\bibfnamefont{D.}~\bibnamefont{Salart}},
  \bibinfo{journal}{Nature} \textbf{\bibinfo{volume}{454}},
  \bibinfo{pages}{861} (\bibinfo{year}{2008}).

\bibitem[{\citenamefont{Reid and Drummond}(1988)}]{reid88}
\bibinfo{author}{\bibfnamefont{M.~D.} \bibnamefont{Reid}} \bibnamefont{and}
  \bibinfo{author}{\bibfnamefont{P.~D.} \bibnamefont{Drummond}},
  \bibinfo{journal}{Phys. Rev. Lett.} \textbf{\bibinfo{volume}{60}},
  \bibinfo{pages}{2731} (\bibinfo{year}{1988}).

\bibitem[{\citenamefont{Reid}(1989)}]{reid89}
\bibinfo{author}{\bibfnamefont{M.~D.} \bibnamefont{Reid}},
  \bibinfo{journal}{Phys. Rev. A} \textbf{\bibinfo{volume}{40}},
  \bibinfo{pages}{913} (\bibinfo{year}{1989}).

\bibitem[{\citenamefont{Reid et~al.}()\citenamefont{Reid, Drummond, Bowen,
  Cavalcanti, Lam, Bachor, Anderson, and Leuchs}}]{reid08}
\bibinfo{author}{\bibfnamefont{M.~D.} \bibnamefont{Reid}},
  \bibinfo{author}{\bibfnamefont{P.~D.} \bibnamefont{Drummond}},
  \bibinfo{author}{\bibfnamefont{W.~P.} \bibnamefont{Bowen}},
  \bibinfo{author}{\bibfnamefont{E.~G.} \bibnamefont{Cavalcanti}},
  \bibinfo{author}{\bibfnamefont{P.~K.} \bibnamefont{Lam}},
  \bibinfo{author}{\bibfnamefont{H.~A.} \bibnamefont{Bachor}},
  \bibinfo{author}{\bibfnamefont{U.~L.} \bibnamefont{Anderson}},
  \bibnamefont{and} \bibinfo{author}{\bibfnamefont{G.}~\bibnamefont{Leuchs}},
  \eprint{arXiv:0806.0270}.

\bibitem[{\citenamefont{Ou et~al.}(1992)\citenamefont{Ou, Pereira, Kimble, and
  Peng}}]{ou92}
\bibinfo{author}{\bibfnamefont{Z.~Y.} \bibnamefont{Ou}},
  \bibinfo{author}{\bibfnamefont{S.~F.} \bibnamefont{Pereira}},
  \bibinfo{author}{\bibfnamefont{H.~J.} \bibnamefont{Kimble}},
  \bibnamefont{and} \bibinfo{author}{\bibfnamefont{K.~C.} \bibnamefont{Peng}},
  \bibinfo{journal}{Phys. Rev. Lett.} \textbf{\bibinfo{volume}{68}},
  \bibinfo{pages}{3663} (\bibinfo{year}{1992}).

\bibitem[{\citenamefont{Silberhorn et~al.}(2001)\citenamefont{Silberhorn, Lam,
  Wei\ss{}, K\"onig, Korolkova, and Leuchs}}]{silberhorn01}
\bibinfo{author}{\bibfnamefont{C.}~\bibnamefont{Silberhorn}},
  \bibinfo{author}{\bibfnamefont{P.~K.} \bibnamefont{Lam}},
  \bibinfo{author}{\bibfnamefont{O.}~\bibnamefont{Wei\ss{}}},
  \bibinfo{author}{\bibfnamefont{F.}~\bibnamefont{K\"onig}},
  \bibinfo{author}{\bibfnamefont{N.}~\bibnamefont{Korolkova}},
  \bibnamefont{and} \bibinfo{author}{\bibfnamefont{G.}~\bibnamefont{Leuchs}},
  \bibinfo{journal}{Phys. Rev. Lett.} \textbf{\bibinfo{volume}{86}},
  \bibinfo{pages}{4267} (\bibinfo{year}{2001}).

\bibitem[{\citenamefont{Schori et~al.}(2002)\citenamefont{Schori, S\o{}rensen,
  and Polzik}}]{schori02}
\bibinfo{author}{\bibfnamefont{C.}~\bibnamefont{Schori}},
  \bibinfo{author}{\bibfnamefont{J.~L.} \bibnamefont{S\o{}rensen}},
  \bibnamefont{and} \bibinfo{author}{\bibfnamefont{E.~S.}
  \bibnamefont{Polzik}}, \bibinfo{journal}{Phys. Rev. A}
  \textbf{\bibinfo{volume}{66}}, \bibinfo{pages}{033802}
  (\bibinfo{year}{2002}).

\bibitem[{\citenamefont{Bowen et~al.}(2003)\citenamefont{Bowen, Schnabel, Lam,
  and Ralph}}]{bowen03}
\bibinfo{author}{\bibfnamefont{W.~P.} \bibnamefont{Bowen}},
  \bibinfo{author}{\bibfnamefont{R.}~\bibnamefont{Schnabel}},
  \bibinfo{author}{\bibfnamefont{P.~K.} \bibnamefont{Lam}}, \bibnamefont{and}
  \bibinfo{author}{\bibfnamefont{T.~C.} \bibnamefont{Ralph}},
  \bibinfo{journal}{Phys. Rev. Lett.} \textbf{\bibinfo{volume}{90}},
  \bibinfo{pages}{043601} (\bibinfo{year}{2003}).

\bibitem[{\citenamefont{Marino et~al.}(2009)\citenamefont{Marino, Pooser,
  Boyer, and Lett}}]{marino09}
\bibinfo{author}{\bibfnamefont{A.~M.} \bibnamefont{Marino}},
  \bibinfo{author}{\bibfnamefont{R.~C.} \bibnamefont{Pooser}},
  \bibinfo{author}{\bibfnamefont{V.}~\bibnamefont{Boyer}}, \bibnamefont{and}
  \bibinfo{author}{\bibfnamefont{P.~D.} \bibnamefont{Lett}},
  \bibinfo{journal}{Nature} \textbf{\bibinfo{volume}{457}},
  \bibinfo{pages}{859} (\bibinfo{year}{2009}).

\bibitem[{\citenamefont{Bowen et~al.}(2002)\citenamefont{Bowen, Treps,
  Schnabel, and Lam}}]{bowen02}
\bibinfo{author}{\bibfnamefont{W.~P.} \bibnamefont{Bowen}},
  \bibinfo{author}{\bibfnamefont{N.}~\bibnamefont{Treps}},
  \bibinfo{author}{\bibfnamefont{R.}~\bibnamefont{Schnabel}}, \bibnamefont{and}
  \bibinfo{author}{\bibfnamefont{P.~K.} \bibnamefont{Lam}},
  \bibinfo{journal}{Phys. Rev. Lett.} \textbf{\bibinfo{volume}{89}},
  \bibinfo{pages}{253601} (\bibinfo{year}{2002}).

\bibitem[{\citenamefont{Howell et~al.}(2004)\citenamefont{Howell, Bennink,
  Bentley, and Boyd}}]{howell04}
\bibinfo{author}{\bibfnamefont{J.~C.} \bibnamefont{Howell}},
  \bibinfo{author}{\bibfnamefont{R.~S.} \bibnamefont{Bennink}},
  \bibinfo{author}{\bibfnamefont{S.~J.} \bibnamefont{Bentley}},
  \bibnamefont{and} \bibinfo{author}{\bibfnamefont{R.~W.} \bibnamefont{Boyd}},
  \bibinfo{journal}{Phys. Rev. Lett.} \textbf{\bibinfo{volume}{92}},
  \bibinfo{pages}{210403} (\bibinfo{year}{2004}).

\bibitem[{\citenamefont{D'Angelo et~al.}(2004)\citenamefont{D'Angelo, Kim,
  Kulik, and Shih}}]{dangelo04}
\bibinfo{author}{\bibfnamefont{M.}~\bibnamefont{D'Angelo}},
  \bibinfo{author}{\bibfnamefont{Y.-H.} \bibnamefont{Kim}},
  \bibinfo{author}{\bibfnamefont{S.~P.} \bibnamefont{Kulik}}, \bibnamefont{and}
  \bibinfo{author}{\bibfnamefont{Y.}~\bibnamefont{Shih}},
  \bibinfo{journal}{Phys. Rev. Lett.} \textbf{\bibinfo{volume}{92}},
  \bibinfo{pages}{233601} (\bibinfo{year}{2004}).

\bibitem[{\citenamefont{Tasca et~al.}(2009)\citenamefont{Tasca, Walborn,
  Ribeiro, Toscano, and Pellat-Finet}}]{tasca09}
\bibinfo{author}{\bibfnamefont{D.~S.} \bibnamefont{Tasca}},
  \bibinfo{author}{\bibfnamefont{S.~P.} \bibnamefont{Walborn}},
  \bibinfo{author}{\bibfnamefont{P.~H.~S.} \bibnamefont{Ribeiro}},
  \bibinfo{author}{\bibfnamefont{F.}~\bibnamefont{Toscano}}, \bibnamefont{and}
  \bibinfo{author}{\bibfnamefont{P.}~\bibnamefont{Pellat-Finet}},
  \bibinfo{journal}{Physical Review A} \textbf{\bibinfo{volume}{79}},
  \bibinfo{eid}{033801} (\bibinfo{year}{2009}).

\bibitem[{\citenamefont{Cavalcanti and Reid}(2007)}]{cavalcanti07}
\bibinfo{author}{\bibfnamefont{E.~G.} \bibnamefont{Cavalcanti}}
  \bibnamefont{and} \bibinfo{author}{\bibfnamefont{M.~D.} \bibnamefont{Reid}},
  \bibinfo{journal}{J. Mod. Opt.} \textbf{\bibinfo{volume}{54}},
  \bibinfo{pages}{2373} (\bibinfo{year}{2007}).

\bibitem[{\citenamefont{Bialynicki-Birula and Mycielski}(1975)}]{bialynicki75}
\bibinfo{author}{\bibfnamefont{I.}~\bibnamefont{Bialynicki-Birula}}
  \bibnamefont{and}
  \bibinfo{author}{\bibfnamefont{J.}~\bibnamefont{Mycielski}},
  \bibinfo{journal}{Commun. Math. Phys.} \textbf{\bibinfo{volume}{44}},
  \bibinfo{pages}{129} (\bibinfo{year}{1975}).

\bibitem[{\citenamefont{Cover and Thomas}(2006)}]{cover}
\bibinfo{author}{\bibnamefont{Cover}} \bibnamefont{and}
  \bibinfo{author}{\bibnamefont{Thomas}}, \emph{\bibinfo{title}{Elements of
  Information Theory}} (\bibinfo{publisher}{John Wiley and Sons},
  \bibinfo{year}{2006}).

\bibitem[{\citenamefont{Monken et~al.}(1998)\citenamefont{Monken, Ribeiro, and
  P\'adua}}]{monken98}
\bibinfo{author}{\bibfnamefont{C.~H.} \bibnamefont{Monken}},
  \bibinfo{author}{\bibfnamefont{P.~S.} \bibnamefont{Ribeiro}},
  \bibnamefont{and} \bibinfo{author}{\bibfnamefont{S.}~\bibnamefont{P\'adua}},
  \bibinfo{journal}{Phys. Rev. A.} \textbf{\bibinfo{volume}{57}},
  \bibinfo{pages}{3123} (\bibinfo{year}{1998}).

\bibitem[{\citenamefont{Walborn et~al.}(2003)\citenamefont{Walborn,
  de~Oliveira, P\'adua, and Monken}}]{walborn03a}
\bibinfo{author}{\bibfnamefont{S.~P.} \bibnamefont{Walborn}},
  \bibinfo{author}{\bibfnamefont{A.~N.} \bibnamefont{de~Oliveira}},
  \bibinfo{author}{\bibfnamefont{S.}~\bibnamefont{P\'adua}}, \bibnamefont{and}
  \bibinfo{author}{\bibfnamefont{C.~H.} \bibnamefont{Monken}},
  \bibinfo{journal}{Phys. Rev. Lett} \textbf{\bibinfo{volume}{90}},
  \bibinfo{pages}{143601} (\bibinfo{year}{2003}).

\bibitem[{\citenamefont{Nogueira et~al.}(2004)\citenamefont{Nogueira, Walborn,
  P\'adua, and Monken}}]{nogueira04}
\bibinfo{author}{\bibfnamefont{W.~A.~T.} \bibnamefont{Nogueira}},
  \bibinfo{author}{\bibfnamefont{S.~P.} \bibnamefont{Walborn}},
  \bibinfo{author}{\bibfnamefont{S.}~\bibnamefont{P\'adua}}, \bibnamefont{and}
  \bibinfo{author}{\bibfnamefont{C.~H.} \bibnamefont{Monken}},
  \bibinfo{journal}{Phys. Rev. Lett.} \textbf{\bibinfo{volume}{92}},
  \bibinfo{pages}{043602} (\bibinfo{year}{2004}).

\bibitem[{\citenamefont{Gisin et~al.}(2002)\citenamefont{Gisin, Ribordy,
  Tittel, and Zbinden}}]{gisin02}
\bibinfo{author}{\bibfnamefont{N.}~\bibnamefont{Gisin}},
  \bibinfo{author}{\bibfnamefont{G.}~\bibnamefont{Ribordy}},
  \bibinfo{author}{\bibfnamefont{W.}~\bibnamefont{Tittel}}, \bibnamefont{and}
  \bibinfo{author}{\bibfnamefont{H.}~\bibnamefont{Zbinden}},
  \bibinfo{journal}{Rev. Mod. Phys.} \textbf{\bibinfo{volume}{74}},
  \bibinfo{pages}{145} (\bibinfo{year}{2002}).

\bibitem[{\citenamefont{Reid}(2000)}]{reid00}
\bibinfo{author}{\bibfnamefont{M.~D.} \bibnamefont{Reid}},
  \bibinfo{journal}{Phys. Rev. A} \textbf{\bibinfo{volume}{62}},
  \bibinfo{pages}{062308} (\bibinfo{year}{2000}).

\bibitem[{\citenamefont{Grosshans and Cerf}(2004)}]{grosshans04}
\bibinfo{author}{\bibfnamefont{F.}~\bibnamefont{Grosshans}} \bibnamefont{and}
  \bibinfo{author}{\bibfnamefont{N.~J.} \bibnamefont{Cerf}},
  \bibinfo{journal}{Phys. Rev. Lett.} \textbf{\bibinfo{volume}{92}},
  \bibinfo{eid}{047905} (\bibinfo{year}{2004}).

\bibitem[{\citenamefont{Gisin}(1991)}]{gisin91}
\bibinfo{author}{\bibfnamefont{N.}~\bibnamefont{Gisin}},
  \bibinfo{journal}{Phys. Lett. A} \textbf{\bibinfo{volume}{154}},
  \bibinfo{pages}{201} (\bibinfo{year}{1991}).

\bibitem[{\citenamefont{Gisin and Peres}(1992)}]{gisin92}
\bibinfo{author}{\bibfnamefont{N.}~\bibnamefont{Gisin}} \bibnamefont{and}
  \bibinfo{author}{\bibfnamefont{A.}~\bibnamefont{Peres}},
  \bibinfo{journal}{Phys. Lett. A} \textbf{\bibinfo{volume}{162}},
  \bibinfo{pages}{15} (\bibinfo{year}{1992}).

\bibitem[{\citenamefont{Bialynicki-Birula}(2006)}]{bialynicki06}
\bibinfo{author}{\bibfnamefont{I.}~\bibnamefont{Bialynicki-Birula}},
  \bibinfo{journal}{Phys. Rev. A} \textbf{\bibinfo{volume}{74}},
  \bibinfo{eid}{052101} (\bibinfo{year}{2006}).

\bibitem[{\citenamefont{Wilk and Wlodarczyk}(2008)}]{wilk08}
\bibinfo{author}{\bibfnamefont{G.}~\bibnamefont{Wilk}} \bibnamefont{and}
  \bibinfo{author}{\bibfnamefont{Z.}~\bibnamefont{Wlodarczyk}}
  (\bibinfo{year}{2008}), \eprint{arXiv:0806.1660}.

\end{thebibliography}


\end{document}